\documentclass{openjournal}
\usepackage{graphicx,amsmath,amssymb,amstext}
\usepackage{amsbsy,amsfonts,amsthm,color,bm}
\usepackage[colorlinks,linkcolor=blue,citecolor=blue,urlcolor=blue ]{hyperref}
\usepackage{cleveref}
\usepackage[utf8]{inputenc}
\setcitestyle{numbers,square}
\usepackage[caption=false]{subfig}
\usepackage{float}
\DeclareMathOperator{\sinc}{sinc}
\begin{document}

\title{Can Einstein (rings) surf Gravitational Waves?}
\author{Leonardo Giani$^*$}
\author{Cullan Howlett}
\author{Tamara M.Davis}

\email{$^*$uqlgiani@uq.edu.au}

\affiliation{The University of Queensland, School of Mathematics and Physics,\\ QLD 4072, Australia}

\begin{abstract}
How does the appearance of a strongly lensed system change if a gravitational wave is produced by the lens? In this work we address this question by considering a supermassive black hole binary at the center of the lens emitting gravitational waves propagating either colinearly or orthogonally to the line of sight. Specializing to an Einstein ring configuration (where the source, the lens and the observer are aligned), we show that the gravitational wave induces changes on the ring's angular size and on the optical path of photons. The changes are the same for a given pair of antipodal points on the ring, but maximally different for any pair separated by $90^{\circ}$. For realistic lenses and binaries, we find that the change in the angular size of the Einstein ring is dozens of orders of magnitude smaller than the precision of current experiments. On the other hand, the difference in the optical path induced on a photon by a gravitational wave propagating \textit{orthogonally} to the line of sight triggers, at peak strain, time delays in the range  $\sim 0.01 - 1$ seconds, making the chance of their detection (and thus the use of Einstein rings as gravitational wave detectors) less hopeless.
\end{abstract}


\section{Introduction}
\label{Introduction}
Einstein rings are spectacular distortions of the images from distant sources induced by the gravitational field of massive structures on the line of sight between them and the observer. The underlying physical process behind these distortions is called \textit{gravitational lensing}, which describes how the trajectories of photons are deformed in presence of a gravitational field. Whilst the phenomena can be qualitatively understood within Newtonian gravity \cite{Soldner}, experiments performed over the last 100 years have shown that General Relativity (GR) is required to correctly describe its physics \cite{Einsteinpred,DysonEDdington,Gilmore,Shapiro,VONKLUBER196047,longo}.

From an historical point of view, gravitational lensing has been indeed one of the first detectable predictions, and thus pieces of observational evidence, in favour of Einstein's theory. Another, recently proven, prediction is the existence of gravitational radiation emitted by time-varying quadrupolar sources of Energy-Momentum. It is straightforward to show that the Einstein field equations, linearized around a Minkowski background metric, become wave equations whose solutions are dubbed \textit{Gravitational Waves} (GW). Indirect observational evidence of these was found as early back as the 1970s \cite{HulseTaylor,Hulsetaylor2}, but direct detection has been possible only over the last decade thanks to the  effort of thousands of scientists  \cite{LIGOScientific:2016aoc}. Nowadays, GW astronomy is one of the most promising fields towards a better understanding of our Universe, both on astrophysical and cosmological scales \cite{KAGRA:2013rdx,Bailes:2021tot}.

The intriguing idea of combining the two aforementioned gravitational phenomena has been subject of in-depth investigations over the last decades. One clear possibility is to study how the propagation of gravitational radiation is affected by the matter distribution along its path. Just like its electromagnetic analog, a gravitational wave may be subject to gravitational lensing, and it is believed that the first strongly lensed gravitational wave will be detected in the near future \cite{Hannuksela:2019kle,Smith:2017jdz,Meena:2019ate,Sereno:2010dr}. On the other hand, another possibility is to study how gravitational waves may influence the path of photons in a strongly lensed system. This topic has been explored in Refs. \cite{PhysRevD.103.103012,Liu:2022fcq,Allen:1990ta,Allen:1989my,Frieman:1994pe}, with focus on determining whether a strongly lensed system may be employed as a detector for very low frequency GWs of cosmological origin. Whilst the answer is in principle yes, it has been shown that an intrinsic degeneracy exists between the lensing configuration and the effects of the gravitational wave, making these two options indistinguishable. However, it seems that such degeneracy does not affect Einstein rings.  The GW may also act as a lens for the propagation of photons, as discussed for example in Refs.~\cite{Damour:1998jm,Faraoni:1997qb}, which concluded that the probability of observing the phenomenon is extremely low. In this work we explore yet another possibility, where a strongly lensed system is perturbed by the propagation of a GW \textit{generated by the lens itself}. Our goal is to quantify the impact of these lens-produced GWs on typical strong lensing observables, such as the angular separation between multiple images and their time delay, and assess the potential of strongly lensed systems as gravitational wave detectors.
The structure of the paper is the following: in section \ref{ER} we briefly review the strong lensing formalism and the configuration needed to produce an Einstein ring.  In Sec. \ref{GWL} we compute the effects of a GW emitted by the lens on the optical path of the lensed photons, under the simplifying assumptions that the GW is linearly polarized and propagating either along or perpendicularly to the line of sight. In Sec. \ref{ORR} we assess the detectability of the aforementioned effects for realistic strong lensing systems.  Finally, sec. \ref{conclusion} is devoted to a summary and discussion of our results.

\section{Einstein rings}\label{ER}
In a strongly lensed system, as long as the geometrical optics description is appropriate, the propagation of light rays from a source $S$ is greatly modified by the gravitational field of a clumpy distribution of matter, the lens $L$, located along the observer's line of sight. 
Whilst the presence of the lens makes it impossible to observe directly the source, the paths of a subset of the source photons (whose initial trajectories would have never reached the observer in the absence of the lens) are distorted in such a way that they eventually intersect the observer position $O$. As a result, the observer will perceive these photons as belonging to spatially different, but otherwise identical sources. The angular separation between the source and its apparent images can be computed with the lens equation, which for a generic mass distribution in the thin lens approximation is
\begin{equation}\label{LE}
    \left( \bm{\beta}- \bm{\alpha} \right) = \bm{\theta}\left(\bm{\beta}\right) \; ,
\end{equation}
where $\bm{\beta} = (\beta_1,\beta_2)$ and $\bm{\alpha} = (\alpha_1,\alpha_2)$ are the position in the sky  of the image and the source respectively measured with respect to the lens position, and $\bm{\theta}\left(\bm{\beta}\right)$ is the reduced deflection angle. We assume that the gravitational system under consideration is described by a perturbed flat FLRW line element 
\begin{equation}
    ds^2 = -\left(1+2\Phi\right)dt^2 +a^2\left(1-2\Phi\right)\left(dr^2 + r^2d\Omega^2\right)\;,
\end{equation}
where $r$ is the comoving radial distance and we have introduced a scalar perturbation $\Phi$ sourced by the mass distribution of the lens, which we assume does not produce any anisotropic stress. The above choice is also known as Newtonian gauge, since at first order in linear perturbation theory for non-relativistic matter the scalar perturbation $\Phi$ satisfies the Poisson equation, just like the gravitational potential in Newtonian gravity.  
The reduced deflection angle for this metric can be written \cite{Bartelmann:2010fz}
\begin{equation}
    \theta_l\left(\beta_m\right) = 2\int_0^{r_s} dr\; \frac{r_s - r}{r}\partial_l\Phi\left(r \beta_m, r\right)\;,
\end{equation}
where $r_s$ is the comoving radial distance to the source.
We can rewrite the reduced deflection angle as the angular gradient\footnote{Derivatives with respect angular coordinates $\beta_i$ on the sky at comoving distance $r$  and partial derivatives with respect to comoving Cartesian
coordinates are related by (see Eq. (50) of Ref.~\cite{Bartelmann:2010fz}) 
\begin{equation}
    \partial_i = \frac{1}{r}\partial_{\beta_i}\;.
\end{equation}} of a scalar function,  $\bm{\theta}\left(\bm{\beta} \right)= \nabla\psi\left(\bm{\beta}\right)$, called lensing potential
\begin{equation}\label{LP}
    \psi(\bm{\beta}) \equiv \frac{2}{c^2 }\frac{\mathcal{D}_{LS}}{\mathcal{D}_L\mathcal{D}_S}\int_{\bm{\beta}}  d\lambda \; \Phi \;.
\end{equation}
In the above equation $c$ is the speed of light, and $\Phi$ is the scalar gravitational potential in the Newtonian gauge, which is integrated along the path of the light ray parametrized by $\lambda$, and depends on the angle $\bm{\beta}$. 
The $\mathcal{D}_i$'s are angular diameter distances, defined as
\begin{equation}
    \mathcal{D}_i(z_i) = \frac{c}{1+z_{i}}\int_0^{z_i} \frac{dz}{H(z)}\;,
\end{equation}
where $z_i$ is the cosmological redshift, $H(z)$ is the Hubble parameter, and $\mathcal{D}_{LS}$ is the angular diameter distance between the source and the lens, $\mathcal{D}_{LS} = \mathcal{D}_S - \mathcal{D}_L$.

\begin{figure}[h]
\includegraphics[scale=0.6,trim=30mm 30mm 0mm 30mm, clip]{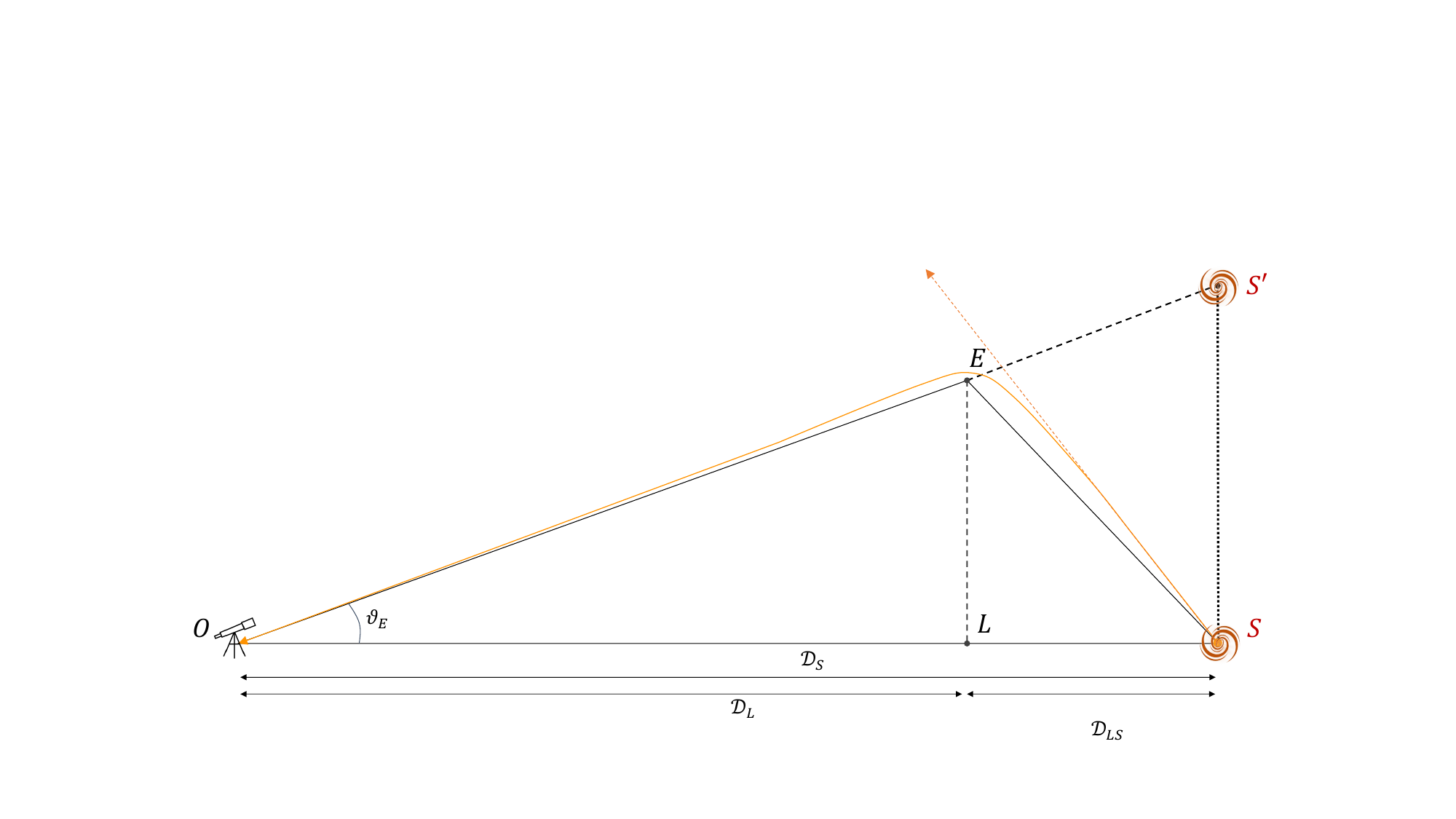}\;
	\caption{A schematic representation of the strong lensing configuration leading to an Einstein Ring. The observer $O$, the source $S$ (at angular diameter distance $\mathcal{D}_S$) and the lens $L$ (at distance $\mathcal{D}_L$) are aligned, with $\mathcal{D}_{LS}$ being the distance between the lens and the source. The yellow solid line represents the trajectory of a photon emitted by the source $S$, bent by the gravitational field of the lens $L$ in such a way that it will eventually meet the observer at $O$. The trajectory that the photon would have without the lens is represented by the dotted yellow line. Because of the bending, the source will appear to the observer at the position $S'$, separated from its real location by an angle $\theta_E$.           }
	\label{Lensingscheme}
\end{figure}

Taking the divergence of Eq.~\eqref{LE}, as long as the extent of the lens is small compared to cosmological distances, we can use the Poisson equation to relate the Laplacian of the lensing potential to the mass distribution of the lens
\begin{equation}\label{LapPsi}
\nabla^2\psi\left(\bm{\beta}\right) = \frac{8\pi G_N}{c^2}\frac{\mathcal{D}_{LS}}{\mathcal{D}_L\mathcal{D}_S} \Sigma(\bm{\beta}) \; ,  
\end{equation}
where $G_N$ is the Newtonian gravitational constant, and we have defined the surface mass density,
\begin{equation}\label{SMD}
    \Sigma (\bm{\beta}) \equiv \int_{\bm{\beta}} d\lambda  \;\rho_L \;,
\end{equation}
which depends on the mass distribution of the lens $\rho_L$. For the assumptions behind Eqs.~\eqref{LE}, \eqref{LP}, \eqref{LapPsi} and their derivation see, for example, Ref.~\cite{Bartelmann:2016dvf}.
Due to their modified paths, the arrival time at the observer position of the photons is delayed with respect to the one they would have in absence of the lens. Furthermore, different images will experience different delays, and we can define 
the time delay between two multiple images $i,j$ as ~\cite{Suyu:2012aa}
\begin{equation}\label{TD}
    \Delta_{ij} = \frac{D_{\Delta_t}}{c}\left( \frac{\left(\bm{\beta}_i - \bm{\alpha}\right)^2}{2} - \frac{\left(\bm{\beta}_j - \bm{\alpha}\right)^2}{2} + \psi\left(\bm{\beta_j}\right) - \psi\left(\bm{\beta_i}\right)\right)\; ,
\end{equation}
where the `time-delay distance' is defined
\begin{equation}\label{timedelaydistance}
    D_{\Delta_t} \equiv \left(1+ z_L\right)\frac{\mathcal{D}_L \mathcal{D}_S}{\mathcal{D}_{LS}} \;.
\end{equation}
It may happen that the gravitational field of the lens is symmetric on the lens plane, and the observer source and lens are perfectly aligned, i.e. $\bm{\alpha}=0$. In this case, instead of having multiple images of the same source, these will appear as a ring-shaped distribution of light, called an \textit{Einstein Ring}. A diagrammatic description of the photons trajectories is given in Fig.~\ref{Lensingscheme}. The angular size $\theta_E$ of the ring is easily computed from Eq. \eqref{LE}, and it is straightforward to realize that the time delay $\Delta_{ij}$ vanishes for any  two points $i,j$ on it since, by definition, $\bm{\beta}_i=\bm{\beta}_j=\theta_E$.

\section{Gravitational Waves produced by the lens}\label{GWL}

We are interested in the effects that gravitational radiation emitted by the lens may have on the trajectories of strongly lensed photons. Gravitational waves perturb geodesics inducing periodic fluctuations in the spatial separation between test particles on the planes perpendicular to the wave's direction of propagation. GWs have in general two polarizations, and are produced whenever a mass distribution has a time-varying quadrupole moment \cite{Maggiore:2007ulw}. A binary system of two massive bodies orbiting around each other, provided that the system is not spherically or rotationally symmetric, will emit gravitational radiation. Gravitational waves detected so far, see Ref.~\cite{GWTC-3} for an updated catalog, have been produced by the inspiral and merging phase of binary systems composed of either stellar mass black holes, neutron stars, or both. Their typical frequencies and scalar amplitudes (\textit{strain}) on Earth are of the order $1-100$ Hz and $10^{-21}$ respectively. 

As a result of the hierarchical formation of massive galaxies, supermassive black holes (SMBHs) are also expected to be found in binary systems, and hence are  potential sources of yet-to-be-detected gravitational waves. However, the typical wavelength of the emitted radiation is orders of magnitude beyond the sensitivity of current ground-based interferometers, and the most promising avenue for detecting it is looking for correlation signatures between the pulse arrival times for a set of known pulsars, known as Pulsar Timing Arrays (PTA) \cite{2010CQGra..27h4013H,Perera:2019sca}.
PTA searches can detect signals in the nanohertz band, where the superposition of radiation emitted by the population of SMBH binaries is expected to result in a stochastic GW background, see Ref.~\cite{2021Symm...13.2418M} for a recent review. On the other hand, sufficiently bright sources can stand above the stochastic signal and therefore be resolved individually. These sources are expected to emit almost monochromatic continuous gravitational waves for decades, with periods ranging from months to years. Up to date, no individual sources have been resolved by the PTA collaborations, and a 95\% sky averaged upper limit on the amplitude of the GW of $h_{95}=9.1\times 10^{-15}$ has been reported in Ref.\cite{IPTA:2023ero}, improving on previous works ~\cite{Babak:2015lua,Schutz:2015pza,NANOGrav:2023bts}.

In a typical strong lensing system the lens is a small and dense cluster of galaxies, which is therefore a suitable candidate to contain a binary system of supermassive black holes. The strain $h$ and the frequency $f_{\rm ISCO}$ of the gravitational wave at the last inner stable circular orbit, felt at an angular distance $\mathcal{D}_{z_i}$ from the binary, are (see  for example Sec. 16.2.2 of \cite{Maggiore:2018sht})
\begin{equation}
    f_{\rm ISCO} = 4.7\left[\frac{\left(m_1+m_2\right)\left(1+z\right)}{10^{3}M_{\odot}}\right]^{-1}\;,
\end{equation}
\begin{equation}\label{straineq}
    h(f,z_i) = \frac{8}{\sqrt{10}}\frac{\left(G\mathcal{M}\right)^{\frac{5}{3}}}{c^4}\frac{\left(\frac{2\pi f_{\rm ISCO}}{1+z_i}\right)^{\frac{2}{3}}}{(1+z_{i})^{2}\mathcal{D}_{z_i}}\;,
\end{equation}
where $\mathcal{M}$ is the Chirp mass
\begin{equation}
    \mathcal{M}=\frac {(m_1m_2)^{3/5}}{(m_1+m_2)^{1/5}}\;,
\end{equation} 
and $m_1,m_2$ are the masses of the black holes.

In the following, we will assume that the lens hosts a SMBH binary, and (for simplicity) that the radiation emitted is linearly polarized. We will consider two different configurations, one in which the GW propagates along the line of sight, and one where it propagates on the Lens plane, perpendicular to the line of sight, as depicted in Figs.~\ref{Fig1} and~\ref{Fig2} respectively. It is important to bear in mind that (to a very large degree) the light rays and the gravitational waves propagate with the same speed \cite{LIGOScientific:2017zic}. As a result, a single photon moving through a portion of space influenced by the gravitational wave will perceive  a \textit{constant} amplitude $h(t_i)$ for the strain, where $t_i$ is the time at which the photon's trajectory was first affected, and \textit{not the strain we would observe on Earth for the same gravitational wave at the arrival time $t_f$}. Therefore, different photons influenced by the GW at different times will travel along different optical paths to the observer.

\subsection{GWs propagating along the line of sight}
\begin{figure}[h]
\includegraphics[scale=0.6,trim=0mm 00mm 0mm 0mm]{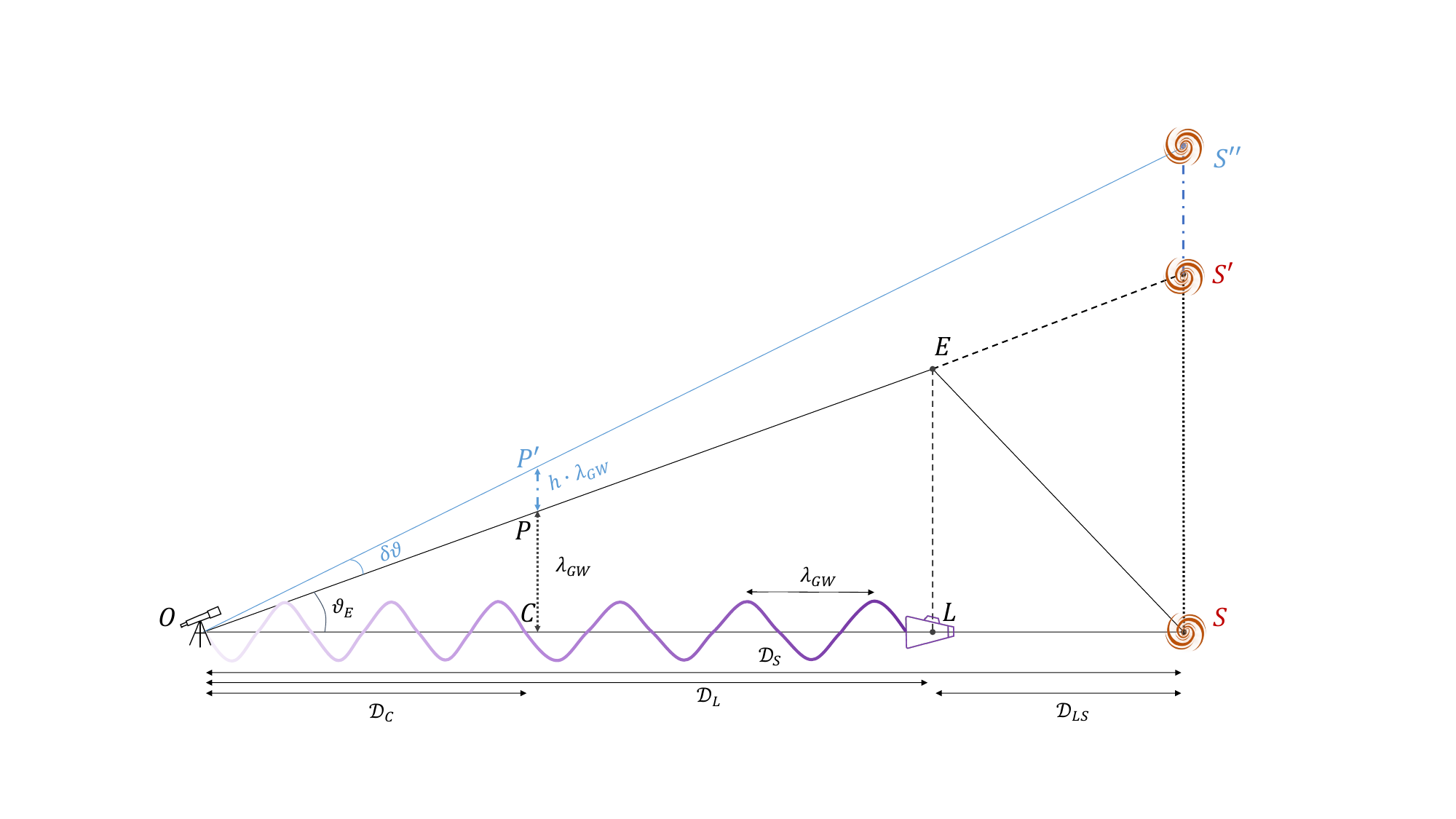}\;
	\caption{A linearly polarized gravitational wave (in purple) is emitted by the lens $L$ and propagates along the line of sight to the observer $O$. A photon traveling from $E$ to $O$ will eventually propagate into a space-time region  perturbed by the gravitational wave, where the spatial separation between two points is stretched or compressed of a factor $(1+h)$ in the directions perpendicular to the line of sight. This happens when the distance between the photon and the line of sight $PC$ is comparable with the wavelength $\lambda_{GW}$ of the gravitational wave. From the point of view of the observer, this is equivalent to observing a photon coming from $P'$, so that the source will appear to the observer at a position $S''$ on the sky, separated from  $S'$ by an angle $\delta \theta$.\\      }
	\label{Fig1}
\end{figure}

Let us consider a GW propagating along the line of sight from the lens to the observer as depicted in Fig.~\ref{Fig1}. The gravitational wave will distort spatial displacements on the plane orthogonal to the direction of propagation, perturbing once more the optical path of the lensed photon.

To assess quantitatively the impact of these perturbations, let us imagine an observer comoving with the GW emitting a photon $\gamma'$ at a point $C$ perpendicular to the direction of propagation of the GW at a time $t_0$. Eventually, this photon will intersect the trajectory of the one emitted by the source $S$ at the point $P'$ at a time $t_1$. A simple calculation in the TT gauge (see for example section 9.1.1 of Ref.~\cite{Maggiore:2007ulw}) shows that the distance traveled by the photon $\gamma'$ can be written

\begin{equation}\label{Maggiorestretch}
\overline{CP'}(t) \approx \overline{CP} \left( 1-\frac{h_{P}}{2}\cos{ \left[\left(t + \frac{\overline{CP}}{2c}\right)\frac{2\pi c}{\lambda_{GW}}\right]}\right)\sinc{\left(\frac{\pi \overline{CP}}{\lambda_{GW}}\right)}\;,     
\end{equation}
where an `overline' denotes the distance between two points (i.e., $\overline{CP'}$ is the distance between $C$ and $P'$). $h_{P}$ is the strain of the gravitational wave felt at $P$, which is inversely proportional to the distance from the GW source $\overline{CL}$, while $\lambda_{GW}$ is its wavelength.\footnote{We thank the anonymous referee for pointing out the similarity between this result and the one obtained in Ref.~\cite{PhysRevD.67.022001} to compute the response of space-based detectors to the passage of a GW.} From this expression, since $\sinc{(x)}$ quickly decays to $0$ when $x\gg1$, we can recognise that the distortion $\overline{PP'}$ hence becomes appreciable only when the distance from the gravitaional wave is of the order of the GW wavelength $\overline{CP}\sim\mathcal{O}\left(\lambda_{GW}\right)$ or smaller.\footnote{The global maximum and minimum of  $\sinc{(x)}$ are given by $x=0$ and $x\approx 1.43 \pi$, corresponding to $\sinc{0}=1$ and $\sinc{1.43\pi}\approx-0.22$ respectively. Therefore we will assume that the perturbations induced by the GW are  negligible for  $\overline{CP}\geq 1.43\lambda_{GW}$.}

As such, in what follows we will describe effectively the perturbation induced by the GW as an instantaneous shift of the source-emitted photon's position from $P$ to $P'$. With reference to Fig. \ref{Fig1}, this happens at a distance $\mathcal{D}_C$ from the observer along the line of sight, such that $\mathcal{D}_C\theta_E \approx \overline{PC}\sim \lambda_{GW}$, when $\sinc{(x)}$ in Eq. \eqref{Maggiorestretch} is of order $1$. If $h\geq 0$ ($< 0$), the photon at $E$ will propagate on a region of space which has been stretched (contracted) by an amount $\overline{PP'}$. Simple trigonometric considerations show that the observer would infer a slightly larger angular size for the Einstein ring $\theta_E + \delta\theta$, such that
\begin{equation}
    \frac{\delta\theta}{\theta_E} \approx h\;.
\label{eq:dtheta}
\end{equation}
The change in the optical path of the photon due to its interaction with the GW, $\Delta \gamma$, can then be computed as
\begin{equation}
    \Delta\gamma = OP'-OP \approx  OP\;h^2\;,
\end{equation}
which is clearly a second order quantity since, at first order in the small-angle approximation,  $\sin\theta \approx \tan\theta$ and hence $OP' = OP = \mathcal{D}_C$.
Because of the different optical path the photon will also be redshifted by a factor,
\begin{equation}
    \Delta z \approx \frac{H(z_C)}{c}\Delta\gamma\;,
\end{equation}
where $H(z_C)$ is the Hubble parameter at the redshift corresponding to the distance where the photon starts to interact effectively with the gravitational wave, $D_C$. We will estimate  $\Delta \gamma$ for realistic strong lensing configurations in Sec.~\ref{ORR}

\subsection{GW propagating along the Lens plane}
 A GW propagating perpendicular to the line of sight will stretch and compress the spatial separations between points along the line-of-sight itself. As a result, the distance from the observer to the source changes and becomes $\overline{OS'}=\overline{OS} + \Delta L$, where $\Delta L$ is the stretch or compression induced by the GW.\footnote{Notice that the spatial extension of the lens in the direction of the line of sight should also change because of the gravitational wave, which we however neglect since we are working within the thin lens approximation.} 

Imagine a photon emitted by an observer, comoving with the gravitational wave, at the position of the lens and directed towards the source. We can then use again \eqref{Maggiorestretch} to infer the change in distance travelled by the photon
because of the gravitational wave. Based on this, we can can estimate the change $\Delta L$ to be of order $\Delta L\approx h\lambda$. Thanks to the symmetry of the problem, we can therefore assume that $\mathcal{D}_S \rightarrow \mathcal{D}_S'=\mathcal{D}_S\left(1 + \Delta L/\mathcal{D}_S\right)$, $\mathcal{D}_L \rightarrow \mathcal{D}_L'=\mathcal{D} L\left(1 + \Delta L/2\mathcal{D}_L\right)$, and $\mathcal{D}_{LS} \rightarrow \mathcal{D}_{LS}'=\mathcal{D}_{LS}\left(1 + \Delta L/2\mathcal{D}_{LS}\right)$. We can thus rewrite the lensing equation in the following form
\begin{equation}\label{modlenseq}
    \left(\theta_E +\delta\theta\right)^2 = 4GM\frac{\mathcal{D}_{LS}'}{\mathcal{D}_{L}'\mathcal{D}_{S}'} \;,
\end{equation}
where we have assumed for simplicity a point mass $M$. In Eq. \eqref{modlenseq} we have neglected the impact of the GW on the lensing potential. This is justified because the gravitational potential associated with a propagating GW, felt at a distance $L$, is of order $\simeq L^2\lambda_{GW}^{-2}h$ (see for example Eq. (9.40) of Ref.\cite{Maggiore:2007ulw}). The change induced by the latter on the lensing potential needs to be compared with the (surface) mass of the lens, and the relative change is orders of magnitude smaller than the effect induced by the distortion of the optical path, whose lower bound is of order $h\lambda/\mathcal{D}_L$ (which is $\approx 10^{-19}$ for a lens at  $10^3$ Mpc). 
Eq.~\eqref{modlenseq} reduces, at first order in $\Delta_L$, to
\begin{equation}\label{deltathetaresult}
    \frac{\delta\theta}{\theta_E}\approx \Delta L\left[\frac{\left(\frac{\mathcal{D}_{L}}{\mathcal{D}_{S}}-\frac{\mathcal{D}_{S}}{\mathcal{D}_{L}}\right)}{\mathcal{D}_{L}-\mathcal{D}_{S}}   \right]\;.
\end{equation}
A more realistic lensing profile than a point mass lens gives different, but qualitatively similar results. Using spheres of matter in isothermal equilibrium to model an extended lens we obtain for the Einstein ring angular size (see for example Eq.(9.3.10) of Ref.\cite{Weinberg:2008zzc}):
\begin{equation}
    \theta_E + \delta_\theta = 4\pi \left<v^2\right>\frac{\mathcal{D}_{LS}'}{\mathcal{D}_S'}\;,
\end{equation}
from which we obtain at first order
\begin{equation}\label{deltathetaisoth}
    \frac{\delta\theta}{\theta_E} \approx \frac{\Delta L}{2}\left[\frac{2\frac{\mathcal{D}_L}{\mathcal{D}_S} -1}{\mathcal{D}_S-\mathcal{D}_L}\right]\;,
\end{equation}
which is qualitatively similar to Eq.\eqref{deltathetaresult}. 

Looking at Eq.~\eqref{deltathetaresult} we easily realize that in the limit $\mathcal{D}_{L}\ll \mathcal{D}_{S}$ the term between square brackets on the right hand side reduces to $\approx \mathcal{D}_{L}^{-1}$. When the lens is halfway between the observer and the source, $\mathcal{D}_{L}= \mathcal{D}_{S}/2$, the square bracket term reduces to $ (3/2)\mathcal{D}_{L}^{-1}$.  In the limit $\mathcal{D}_{L}\rightarrow\mathcal{D}_{S}$ the square bracket term is approximately $\approx 1$ (this latter case is quite unrealistic).

\begin{figure}[h]
\includegraphics[scale=0.6,trim=00mm 0mm 0mm 0mm]{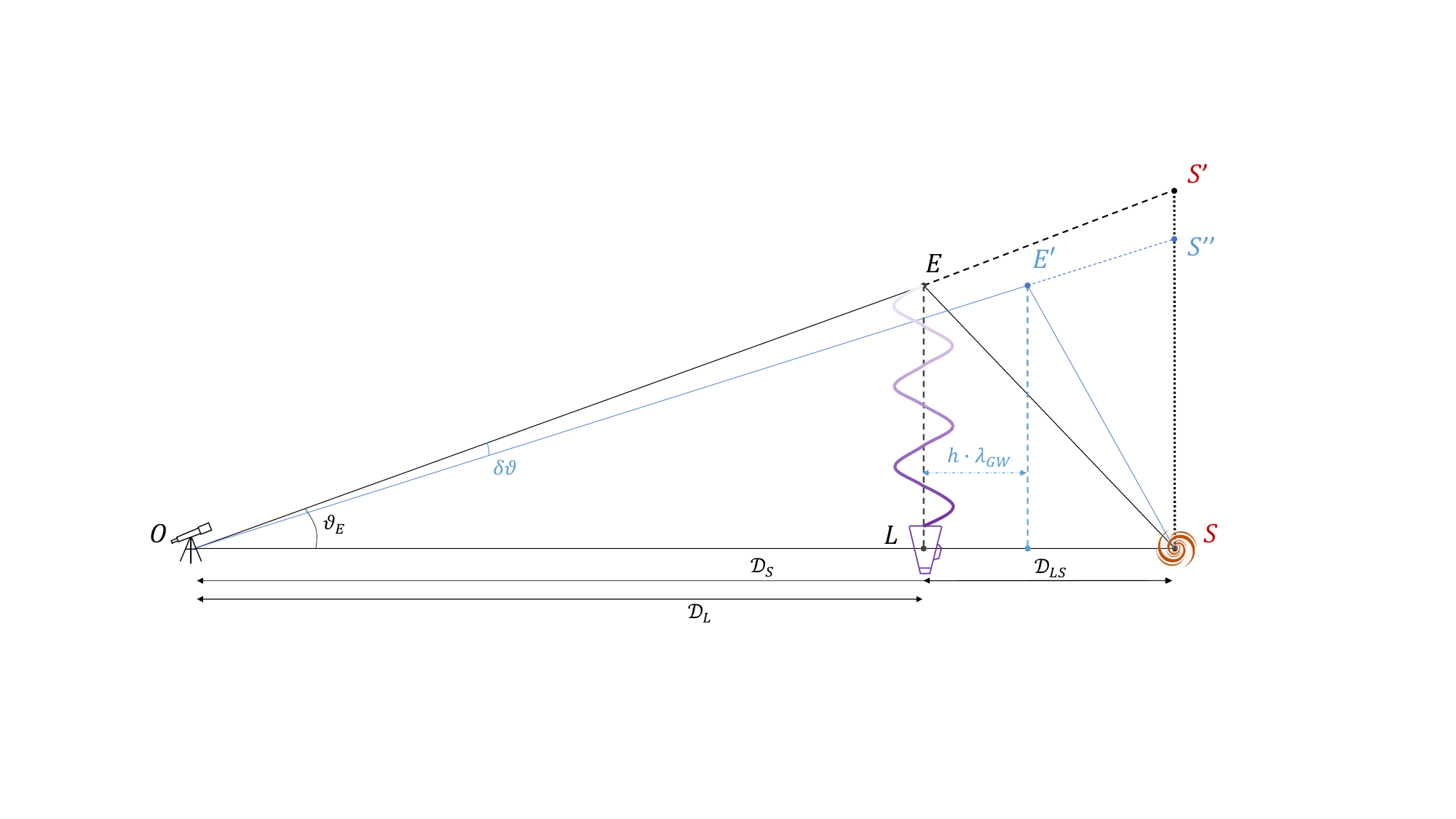}\;
	\caption{A linearly polarized gravitational wave (in purple) is emitted by the lens $L$ and propagates in the lens plane towards $E$. A photon approaching $E$ will eventually propagate into a space-time region perturbed by the gravitational wave, where the spatial separation between two points is stretched of a factor $(1+h)$. This happens when the distance between the photon and $E$ is comparable with the wavelength $\lambda_{GW}$ of the gravitational wave. From the point of view of the observer, this is equivalent to observing a photon coming from $E'$, and  the source will appear at the position $S''$, separated from  $S'$ of an angle $\delta \theta$.}
	\label{Fig2}
\end{figure}

From the perspective of a photon travelling from the source $S$, we conclude that we can effectively describe the phenomenon as a change in the position of the point about which it is deflected. This configuration is depicted in Fig.~\ref{Fig2}, where the distance from the observer to the lens has become $\mathcal{D}_L +\overline{EE'}$.
If we assume that the lens is sufficiently far away from the source, we can roughly estimate the displacement $\overline{EE'}$ to be of order  $\sim$ $h\lambda_{GW}$ (as depicted in Fig. \ref{Fig2}),  with a resulting shift on the angular size of the ring at first order in $h$ given by
\begin{equation}
\frac{\delta\theta}{\theta_E}\approx -h\;\frac{\lambda_{GW}}{\mathcal{D}_L}\;.    
\end{equation}
Note that this differs by a factor $\lambda_{GW}/\mathrm{D_{L}}$ from the result in Eq.~\ref{eq:dtheta} for a GW propagating along the line-of-sight. As we will show in the next section, this factor is typically a very small number but is countered to a small degree by the fact that the strain the photon experiences --- which must be evaluated at a length scale of order $\sim \overline{LE}$ from the binary --- is much larger in this scenario.

The corresponding difference in the optical path of the photon is
\begin{equation}
    \Delta\gamma = OE'-OE \approx -\frac{\mathcal{D}_L}{1+z_L}\frac{\delta \theta}{\theta_E} \;,
\end{equation}
 which induces a redshift on the photon frequency of order
\begin{equation}
     \Delta z \approx \frac{H(z_L)}{c}\Delta\gamma\;.
\end{equation}
We will estimate  $\Delta \gamma$ for realistic strong lensing configurations in the following section.

\section{Realistic scenarios: could ringing rings be observed?}\label{ORR}
To assess the significance of the effects described so far, let us specialize to a reasonably realistic configuration.\footnote{Similar to the one producing the Einstein ring \href{https://hubblesite.org/contents/media/images/2005/32/1794-Image.html}{SDSS J073728.45+321618.5}. } Assume the source is at redshift $z_S \approx 0.6$ and the lens at redshift $z_L \approx 0.3$. 
If the lens contains a binary system of SMBHs with masses $m_1 = m_2 \sim 10^{10} M_{\odot}$, this sources a gravitational wave with frequency of order $
f_{\rm ISCO} \sim 10^{-7} \,\mathrm{Hz}$ and corresponding wavelength of order $\lambda_{GW} \sim 10^{-1} \,\mathrm{pc}$. 
Scaling relations of the ratio between the SMBH masses and the total stellar mass suggest that such a SMBH binary would have a surrounding total stellar mass between $10^{11.5}-10^{12} M_{\odot}$, see for example Fig.~8 of Ref.~\cite{2015ApJ...813...82R} or Fig.~4 of Ref.~\cite{2018ApJ...869..113D}. The total mass, including the DM contribution, can thus be inferred using scaling relation between the total stellar mass and the halo mass (see for example Fig.~9 of Ref.~\cite{Girelli:2020goz} ), and turns out to be of order $\sim 10^{12}-10^{14} M_{\odot}$. 
 
Assuming $M_{L} \approx 10^{12} M_{\odot}$, for the configuration depicted in Fig. \ref{Fig1} we find that at the point $C$ the strain is roughly the same as the strain at the observer's position, $h\sim 10^{-14}$, and hence
\begin{equation}
    \frac{\delta\theta}{\theta_E} \approx 10^{-14}\;, \qquad \Delta\gamma \approx 10^{-4} \,\mathrm{cm}\;, \qquad \Delta z \approx 10^{-32}\;.
\end{equation}
Due to the longer optical path traveled by the photon, we will also observe a delay in its arrival time of order
\begin{equation}
    \Delta t = \Delta\gamma/c \approx 10^{-15} \,\mathrm{s} \;.
\end{equation} 
We consider all of the above to be completely immeasurable.

However, in the configuration depicted in Fig. \ref{Fig2} for the same lens we find instead that the strain at the point $E$ is of order $h\sim 10^{-9}$, and we can compute
\begin{equation}
        \frac{\delta\theta}{\theta_E} \approx 10^{-19}\;, \qquad \Delta\gamma \approx 10^{9} \,\mathrm{cm}\;, \qquad \Delta z \approx 10^{-19}\;.
\end{equation}
Notice that in this case the photon trajectories intersect the gravitational wave at distances from the binary system $\approx 10$ orders of magnitude smaller than in the previous case, where the strain is 5 orders of magnitude bigger.  As a result, the optical path difference $\Delta\gamma_B$ increases significantly, and triggers a delay on the time of arrival of the photon of order
\begin{equation}
    \Delta\gamma/c \approx 0.01 \,\mathrm{s} \;.
\end{equation}

\begin{figure}[]
\includegraphics[scale=0.92,trim=00mm 0mm 0mm 00mm]{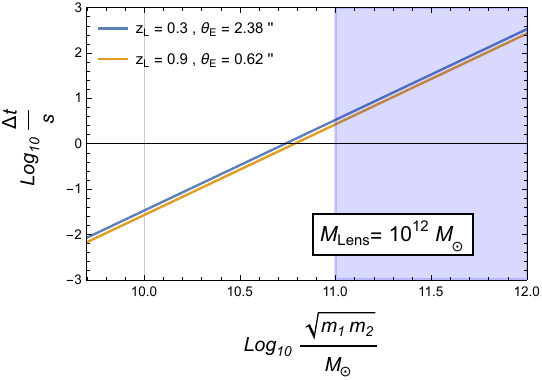}\;
\includegraphics[scale=0.92]{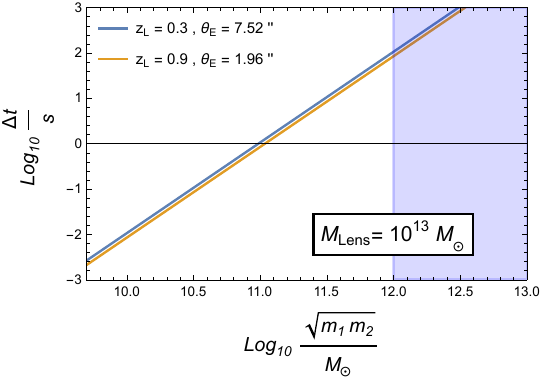}\;
\begin{center}
\includegraphics[scale=0.92,trim=0mm 0mm 0mm 0mm]{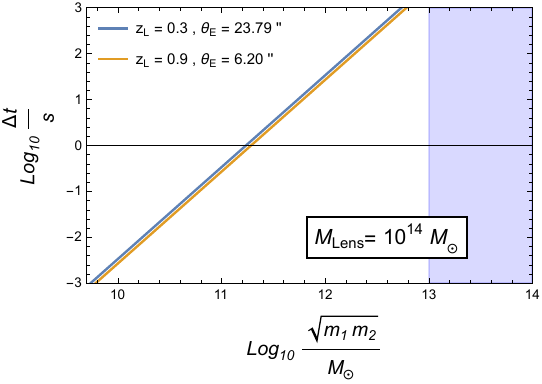}\;
\end{center}
	\caption{Plots of the time delay, in seconds, induced by a gravitational wave from a SMBH binary at peak strain as a function of the product of the black hole masses. The source is at redshift $z_S = 1$, and we considered two different lenses at redshift $z_L= 0.3$ and $z_L= 0.9$. The corresponding angular size of the Einstein ring is given in arcseconds. In the blue region the masses of the black holes constitute $ >10\%$ of the total mass of the lens, which we consider to be unrealistic.}
	\label{Niceplot}
\end{figure}

Being the most significant of the effects discussed so far, we studied how the latter prediction changes if we consider  different lensing configurations or different binary systems. In Fig.~\ref{Niceplot} we show the induced time delay as a function of (the square root of) the product of the two masses $\sqrt{m_1m_2}$ in the binary, where the angles $\theta_E$ entering the calculation were estimated using the lens equation for a point mass lens $M_{Lens}$. The source is at redshift $z=1 $, and we consider two different distances from the observer to the lens and different lens masses. If we fix the masses of the black holes, the plots show that the time delay increases for lighter lenses. If we fix the lens mass instead, the time delay increases with heavier binaries. We found that time delays of the order of $\sim 1\mathrm{s}$ can only be reached for black holes with masses $m_{1},m_2 \sim 10^{11} M_\odot$ and lenses with mass $M_{Lens} \approx 10^{12}-10^{13} M_{\odot}$. Even though SMBH with such masses might exist, see for example Ref.\cite{2016A&A...585A.153B}, their properties are not well understood and extrapolating the standard scaling relations for such large masses could be misleading.
Finally, the time delay is inversely proportional to $(1+z_L)$, and hence increases when the lens is closer to the observer. 

\section{Discussion}\label{conclusion}
We computed the impact of the gravitational radiation emanated by a binary system of two supermassive black holes on strong lensing observables in a typical configuration. Since the gravitational wave induces periodic fluctuations in the size of spatial displacements with dimension comparable to or smaller than its wavelength, the main effect on the system is to change the optical path of photons in their journey from the source to the observer. This, consequently,  affects: $i)$ the angular separation of the multiple images of a gravitationally lensed source, $ii)$ their apparent redshift, and $iii)$ their time of arrival at the observer position. 

To get a qualitative understanding of the phenomenon, we considered two different configurations, portrayed in Figs.~\ref{Fig1} and \ref{Fig2}, with the simplifying assumptions that the gravitational wave is linearly polarized and propagates colinearly or orthogonally to the line of sight.
In the former case the ring's shape becomes ellipsoidal, with opposite periodic deformations along the $x$ and $y$ axes. In the latter case the ring is also distorted into an ellipse, but with the $y$ axis fixed and the oscillations occuring only along the $x$ direction. In this case we also observe the overall size (i.e. Einstein radius) of the ring to be oscillating, as a result of the fluctuations along the $z$ direction.


Given the typical sensitivity of current experiments, of order $\sim 0.01''$ for the angular separation of the images \cite{2012A&A...538A..99S} and of the order of $\sim 10^{-6}$ for the redshifts \cite{Carr:2021lcj}, there is no hope of detecting  $i)$ or $ii)$ in either of the configurations of Figs.~\ref{Fig1} or \ref{Fig2}. On the other hand, for the configuration of Fig. \ref{Fig2}, the significance of $iii)$ varies between $\sim10^{-2}-10^{1}\,\mathrm{s}$. The possibility of observing this, on the basis of qualitative orders of magnitude considerations, seems less hopeless. 

Let us briefly put in perspective the challenges involved with such a measurement. Since we are targeting a gravitational wave with frequency $\sim 10^{-7}-10^{-8}\,\mathrm{Hz}$, the duration of the transient signal is of the order of the $\sim 10^{-1}-10^{0}\,\mathrm{yr}$. For an idealized Einstein ring, in the absence of a gravitational wave, there is no time delay between the arrival time of two photons coming from randomly chosen points on the ring. Let us choose two points $(A,B)$ separated by an angle $\pi/2$ on the ring, lying on the principal axes $x $ and $y$ orthogonal to the direction of propagation of the gravitational wave. Because of the latter, lengths $l$ on the $x$ axes are increased of a factor $(1+h)$ whereas on the $y$ axes they are decreased of a factor $(1-h)$. Therefore, the time delay will not vanish anymore and at peak strain, after a few months or a year, we should be able to measure a time delay between the light curves from the points $A,B$ of the order $\sim 10^{-2} \,\mathrm{s}$ for typical SMBH masses, and up to the order of $10^{-1} - 10^{0} \,\mathrm{s}$ for exceptionally massive ones. Of course, to detect a time delay one also needs a time varying source, which adds to the difficulty of finding an appropriate lens system. Typical target sources with intrinsic variability which are used or have been proposed to measure strong lensing time delays are quasars \cite{Bonvin:2015jia,Bonvin:2016crt,Courbin:2010au,Courbin:2017yvz}, double pulsars \cite{Rafikov:2005gc,Lai:2004ue,1996PASA...13..236W} and repeating Fast Radio Bursts \cite{Li:2017mek,Liu:2019jka,Wucknitz:2020spz,Pearson:2020wxb}.

A slightly more optimistic (but less realistic) scenario  would involve two black holes with masses of order $10^{11} M_{\odot}$ constituting a significant fraction ($\geq 10 \%$) of the lens. In this case, the time delay would be of the order of few seconds.  Of course, in this exercise we have made a number of simplifying assumptions which are unlikely to occur in real astrophysical systems, like considering a perfect Einstein ring arising from an idealized lens profile, and the ideal orientation and polarization of the gravitational wave with respect to the line of sight. Relaxing these assumptions should not change the qualitative picture given here, but will likely introduce systematics and reduce the likelihood of any detection. Nevertheless, and surprisingly, the intrinsic magnitude of the effect is appreciable on typical human time-scales. We conclude that, even if difficult, indirect detection of gravitational waves produced by supermassive black holes through long-term monitoring of strongly lensed systems might be feasible in the future.

We would like to stress that direct detection of gravitational radiation from SMBH binaries with $10^{10} M_{\odot}$ is unlikely to be feasible with LISA, whose sensitivity is in the range $10^{-4}-10^{1}$ Hz, and therefore will be able to target binaries with masses within $10^{4}- 10^{7} M_{\odot}$ \cite{Colpi}. On the other hand, PTA searches for GW from a single source can detect radiation produced by SMBH binaries with Chirp mass $M\geq 10^{9} M_{\odot}$ and frequencies between ${10^{-9}-10^{-7}}$ Hz, overlapping with the sources of indirect detection discussed in this paper. According to the analysis of Refs.~\cite{Moore:2014eua} these signals should be detectable with a signal-to-noise ratio (SNR) above a chosen threshold  SNR $\geq 3$, despite the fact that none of these events have been detected so far~\cite{IPTA:2023ero}.

\section*{Acknowledgements}
We are grateful to Paul Lasky, Riccardo Sturani and Oliver F. Piattella for useful comments and suggestions.
The authors acknowledge support from the Australian Government through the Australian Research Council Laureate Fellowship grant FL180100168.



\bibliographystyle{apsrev4-2.bst}

\bibliography{Ref.bib}

\end{document}